# Acoustoelectric current in graphene due to electron deformation potential and piezoelectric phonon couplings

Subhana Nafees, S. S. Z. Ashraf, M. Obaidurrahman
*Department of Physics, Faculty of Science,*
*Aligarh Muslim University, Aligarh-202002, Uttar Pradesh, India*

Email: *ssz_ashraf@rediffmail.com*

**Abstract**

Recent studies strongly indicate that graphene can be used as a channel material for converting surface acoustic waves to acoustoelectric current, which is a resource for various exciting technological applications. On the theoretical side, studies on phonon amplification/attenuation and acoustoelectric current at low temperatures in graphene have reported approximate analytical results under exceedingly simplifying conditions using the Boltzmann transport equation. Overcoming the earlier simplifying assumptions, we investigate both numerically and analytically the governing kinetic equations for amplification/attenuation and acoustoelectric current, taking into account the piezoelectric and deformation potential electron phonon coupling mechanism in the semi classical Boltzmann transport formalism approach, and obtain analytical results that are in reasonable agreement with the reported experimental results.

**Keywords:** Acoustoelectric current, graphene, deformation potential, piezoelectric, electron phonon coupling, Boltzmann Transport formalism

## I. INTRODUCTION

The coupling of acoustic phonons with electrons is a phenomenon that serves both as a source of exciting physics and as a resource for novel device applications that emerge from time to time [1]. Phonons or phonon based mechanisms are an integral part of many modern technologies in which phonons can be manipulated in materials for novel device applications, including the promising functionalities for carrying and processing quantum information [1-4]. This has given rise to a competing fast emerging field of phononics in parallel to electronics and photonics. A significant effect related to the interaction of phonons with electrons in materials discovered in the early '50s is the acoustoelectric effect (AE) that has been of much practical importance. The AE is observed both in bulk and nano specimen where a dc voltage (if open-circuited) or a dc electric current develops in the direction of propagation of the acoustic wave. The AE occurs by transferring some part of momentum by the acoustic wave to the electron gas in the process of electronic absorption [5-6].

Apart from the bulk acoustic waves, surface acoustic waves (SAW) too have been of immense use in diverse novel device applications. SAW has an advantage that it travels along the surface of a material and thus is accessible all along its path of propagation. The SAW modulates the electron density and exchanges energy with the electrons via the Deformation potential (DP) and Piezoelectric (PE) interactions. In case of piezoelectric materials, SAW is accompanied by an electric field through which it can strongly interact with electrons within the adjacent surface and excite electric current in it. Nowadays SAW are induced by comb-like metallic structures, called inter digital transducers (IDTs), deposited on the surface of the piezoelectric substrate. The piezoelectric effect causes the high frequency input signal at the transmitting IDT stimulate the SAW. Applications of SAW are too numerous to quote and have been categorically mentioned in research papers, review articles and monographs [1-8]. A latest application of SAW finds use in life sciences in the process of biosensing, cell monitoring and manipulation, and also in microfluidics for sensing and fluid mixing [9]. A whole new field of high frequency sono-processing for caveat free acoustic material synthesis processing and manipulating is fast emerging [10]. In a recent review article the snapshot of applications and future roadmap of SAW devices has been elaborated [11].

The advent of graphene on the material landscape has enriched the scope of acoustoelectric phenomenon through the occurrence of SAW in this novel material [12].A mono-layer graphene (MLG) is a single two-dimensional hexagonal sheet of carbon atoms which are $sp^2$ hybridized giving rise to three in-plane σ bonds and a π orbital perpendicular to the plane. While the strong σ bonds is the cause of robust rigidity of the graphene sheet, the out-of plane π bonds is behind the enhanced electrical and thermal mobility of the charge carriers. The low energy excitations in MLG are governed by Dirac equation which leads to the observation of unusual properties as compared to metals and semiconductors [13-14]. The richness of electronic, optical, optoelectronics, photonic and thermal properties of graphene because of its high electrical and thermal conductivity, optical transparency, mechanical flexibility and robustness, and environmental stability has established MLG as a material of prime importance for wide range of novel technological applications [15-16].

As pointed above graphene has emerged as a highly promising material for acoustic applications also, because of its inherent capacity to generate large acoustic phonon frequencies of order of 10 THz that is much higher than that observed in bulk GaAs and conventional semiconductors. This has opened up a new possibility for realizing THz phononic devices. The strong sound absorption properties of surface acoustic waves in a wide frequency range up to terahertz at room temperature has the potential to provide a new sustainable solution to reduce the emerging environmental pollution of turbo machinery noise and vibration [17-18]. The zero band gap in MLG limits its use as a switch in Logic circuits but now AE current in MLG can enable overcome this inherent limitation through the construct of graphene based AE transducer that can function as a logic switch [19].The AE of the SAW photo amplification can be utilized in the creation of an opto acousto electric on graphene-piezoelectric structures for collecting, amplifying, and detecting super weak sources of THz-radiation photons [20]. The amplitude of the SAW is affected when target molecules are dropped onto the propagation path of a SAW.

This property is anticipated for use in the simultaneous detection of charge and mass by combining Graphene Field Effect Transistors and SAW sensors [21].

The coupling between charge carriers in thin films placed on a piezoelectric (PE) substrate with SAW is induced mainly by the presence of PE potential accompanying the propagating SAW. In a PE crystal a mechanical strain produces an electric field proportional to the strain, and in a non PE crystal a deformation of the lattice produces a change in potential energy of the conduction electrons proportional to the strain. In Fig.1(a) is depicted the schematics of a Graphene sheet placed on a PE substrate between two IDT devices. A SAW is generated by a high-frequency signal input to an IDT1 formed on a PE substrate. When a SAW propagates in graphene, an AE current ($I_{AE}$) flows between two attached electrodes as shown in Fig.1a. From the existing experimental and theoretical studies exploring the interaction of charge carriers with sound in MLG, it turns out that it is still difficult to completely understand the dynamic properties of AE effects including the $I_{AE}$, as it is influenced by a large number of parameters which includes the properties of graphene [12], intensity and frequency of the SAW [17,20-33], temperature [22], time [24], carrier concentration[17, 29], doping [24], magnitude and sign of bias voltage (V bias) [23,26,29], illumination [20], substrate, as well as type of piezoelectric crystal and the interface between the graphene and the piezoelectric crystal surface [25-26,30].

We present a theoretical investigation on the AE current ($I_{AE}$) generated by acoustic phonons interacting with electrons through piezoelectric (PE) and deformation potential (DP) electron phonon coupling in a MLG using Boltzmann transport equation (BTE) approach. In a few theoretical studies the investigations on attenuation/absorption coefficient/rate and acoustoelectric current in the BTE approach have been reported [30-33]. The results have been obtained in the hyper sound regime where $ql>>1$. The mentioned studies have reported analytical expressions for the said quantities under simplifying assumptions of Fermi Dirac (FD) distribution function which remain valid only at temperatures far below the room temperature, where the approximations undertaken to derive the analytical results are valid. The difference in the FD functions used at lower and higher temperatures has been illustrated in fig.1b, where a plot of ratio of the difference between the complete and approximate FD functions, $f(K) - f(K^{'})/\text{Appr.}(f(K) - f(K^{'}))$ at $\omega_q = 1 THz$ and $k = 2 \times 10^8 m^{-1}$. Clearly the ratio is 1 in the low temperature limit and above $(T > 30K)$ the ratio begins to decline. So the reported results under this approximation at higher temperatures may not represent the accurate picture and hence are suspect. This warrants a complete evaluation of the kinetic equations and the rechecking of the parameter dependencies for phonon amplification and of the AE current, so that a better correspondence with experiments could be established.

The paper is organized as follows. The sec. II describes the formalism using Boltzmann transport equation and the obtained analytical results for electron-DP and PE couplings. In Sec. III, the numerical evaluation of the complete expression and their comparison with analytical results are discussed. Finally in Sec IV the study is concluded.

## II. FORMALISM AND ANALYTICAL RESULTS

The peculiar linear energy dispersion relation of graphene is given by $\varepsilon_{\pm}(\vec{k}) = \pm \hbar v_f |\vec{k}|$ and the charge carrier is described by the two component wavefunction $\Psi_{\pm\vec{k}}(\vec{r}) = e^{i\vec{k}\cdot\vec{r}}(1, \pm e^{i\varphi})/\sqrt{2}$. Here, $\pm$ refers to electronic and hole charge carriers, $v_f$ is the Fermi velocity, $\varphi$ is the angle between $k$ and $\vec{r}$ directions, $\vec{r}=(x,y)$ is the position vector in two dimensions, $\vec{k} = (k_x, k_y)$ is the wavevector for a carrier along the graphene sheet. In tight binding approximation the effective electron phonon Hamiltonian for MLG is written as $H = H_0 + H_{e-p}$ where $H_0$ represents the two components Hamiltonian near the K points that corresponds to free electron and $H_{e-p}$ is the electron phonon interaction and is given as $H_{e-p}^{DP(PE)} = \sum_{kq} M_{kq}^{DP(PE)} a_{k+q}^{\dagger} a_k (b_q + b_{-q}^{\dagger})$, where $M_{kq}^{DP(PE)} = M_{kk'}^{DP(PE)}(q) C_{kk'}$ in which $M_{kk'}^{DP(PE)}$ is the electron-phonon coupling strength for intrinsic DP scattering ( extrinsic PE acoustic scattering), the chirality factor $C_{kk'} = |C_{k+q}^{\dagger} C_k|^2 = (1 + \cos\theta_{kk'})/2$ that arises from the overlap of wave functions, $a_{k+q}^{\dagger}(a_k)$ are the electron creation (annihilation) operators, $b_q(b_{-q}^{\dagger})$ are the phonon annihilation (creation) operators, $\theta$ is the angle between scattering in and out wave vectors $k$ and $k'$. The transition rate induced in the system due to electron phonon interaction is given by Fermi's golden rule, $T_{kk'} = \frac{2\pi}{\hbar} \sum_q |\langle k'|H_{e-p}|k\rangle|^2 C_{KK'} \delta(E_k \pm \hbar\omega_q - E_{k'})$. Considering that the scattering mechanism involves both the absorption and the emission process, the square of the matrix element of the electron-phonon interaction evaluates out as $|\langle k'|H_{e-p}|k\rangle|^2 F_{KK'} \delta(E_k \pm \hbar\omega_q - E_{k'}) = |M_{kq}|^2 C_{KK'} \{N_{\omega_q} \delta(E_k + \hbar\omega_q - E_{k'}) + (N_{\omega_q} + 1)\delta(E_k - \hbar\omega_q - E_{k'})\}$, where $(N_{\omega_q} = 1/\{\exp(\beta\hbar\omega_q) - 1\})$ is the Bose Einstein (BE) distribution function for phonons at temperature $\beta = \frac{1}{K_B T}$ ($T$ is the temperature and $K_B$ is the Boltzmann constant). The delta function $\delta(E_k \pm \hbar\omega_q - E_{k'})$ ensures energy conservation for the the inelastic scattering processes by the absorption and emission of phonons of wavevector $q = k - k' = 2k_F \sin\left(\frac{\theta}{2}\right)$. The square of the matrix element for the transition rate from k to k' for electron acoustic phonon is given as,

$$\left|M_{kk'}^{DP(PE)}(q)\right|^2 C_{kk'} = \begin{cases} \frac{V^{DP^2} \hbar\omega_q}{2A\rho v_s^{DP^2}} \left(\frac{1+\cos\theta}{2}\right) \\ \frac{c_{PE}^2 V^{PE^2} \hbar e^{-2qd}}{2\pi A\rho_s v_s^{PE}} \frac{q_x q_y}{q^2} \left(\frac{1+\cos\theta}{2}\right) \end{cases} \quad (1)$$

Where $V^{DP}(V^{PE})$ is the coupling constant for electron-DP (electron-PE) phonon interaction, $A$ is the area of graphene specimen, $\rho(\rho_s)$ is the graphene areal mass density (substrate mass density), $v_s^{DP}(v_s^{PE})$ is velocity of sound in material for DP(PE) interaction, $c_{PE}$ is a constant arising out of the elastic properties of GaAs substrate, $d$ is the distance between graphene and PE substrate. As can be seen from the above matrix elements for the dispersion relation, $\omega_q = v_s^{DP(PE)} q$, the DP interaction Hamiltonian is proportional to $\sqrt{q}$, whereas in contrast, the PE interaction

Hamiltonian is proportional to $1/\sqrt{k}$. The piezoelectric interaction will therefore be more important at low $k$, that is, at low temperature. On comparing the relative magnitude of PE and DP phonons matrix elements, $|M_{kk'}^{PE}(q)|/|M_{kk'}^{DP}(q)| = 1.3 \times 10^7/q\ m^{-1}$ suggesting that both these scattering mechanisms are important in providing important channels for energy relaxation.

The interaction between a SAW and the two dimensional electron gas in doped graphene layer placed over the substrate induces a linear acousto electric direct current density, $J_{AE}$. This $J_{AE}$ as stated above develops in a closed circuit in the specimen in direction of propagation of acoustic waves due to the transmission of momentum to the electron gas (electronic absorption) by the acoustic phonon population $N_q$ characterizing the acoustic wave. When the system is in equilibrium (i.e., external driving field is absent) the distribution of phonons is given by BE distribution of acoustic phonons at the lattice temperature. The momentum transferred to the electrons is $\frac{d\mathbf{p}}{dt} = -\hbar q \frac{dN_q}{dt}$ where $\frac{dN_q}{dt}$ is the number of phonons per unit time absorbed or emitted by the electrons. The energy transferred to the electrons is $\frac{dE}{dt} = -\hbar\omega \frac{dN_q}{dt}$. The absorption coefficient $\Gamma = -\frac{1}{\Phi}\frac{dE}{dt}$ is just the energy transferred per unit time divided by the incident energy flux $\Phi$. This relation was originally derived by Weinreich and in the absence of external dc electric and magnetic fields [34]. The kinetic equation for the acoustic phonon population $N_q$ in the medium is governed by the following equation [29]

$$\frac{dN_q}{dt} = \frac{1}{\tau_q} N_q \qquad (2)$$

Where $\tau_q$ is the electron-acoustic DP(PE) phonon relaxation time. This time for electron–phonon scattering can vary widely depending on the nature of electron–phonon interaction mechanism and the electron energy distribution. Typical electron–phonon scattering times are in general of the order tens of picoseconds. Denoting, $\tau_q = R_q^{-1}$ where $R_q$ is the phonon amplification/attenuation rate ( $R_q > 0$, amplification occurs and when $R_q < 0$ , absorption or attenuation occurs) , $\frac{\partial N_q(t)}{\partial t} = R_q N_q(t)$ or,

$$R_q = \frac{1}{N_q}\frac{\partial N_q}{\partial t} \quad (3)$$

The total rate of phonon absorption and emission from the perturbation theory can be expressed as,

$$\frac{\partial N_q(t)}{\partial t} = (2\pi/\hbar)\, g_s g_v \sum_{k,k'} \left| M_{kk'}^{DP(PE)}(q) \right|^2 \{(N_q + 1)C_{KK'}\delta_{k,k'} \times f(K)(1 - f(K'))\delta(E_k - E_{k'} - \hbar\omega_q) - N_q f(K')(1 - f(K)) \times \delta_{k,k'-q}\delta(E_k - E_{k'} - \hbar\omega_q)\} \quad (4)$$

In eq.(4) $g_s = 2, g_v = 2$ are the spin and valley degeneracies, respectively and $E_{k'} = E_k + \hbar\omega_q$. The $f(K)$ is the shifted equilibrium electronic FD distribution function in the linear response regime approximated as $f(K) = \{\exp[\beta(\hbar v_f K - \varepsilon_f)] + 1\}^{-1}$, where $K = \sqrt{k^2 - 2kk_d\cos\phi + k_d^2}$ and $k_d = k_f v_d/v_f$ being the shift of the electron momentum due to the presence of the drift velocity, $v_d$ caused by the application of the dc electric field $E_d = V_{sd}/L$ with $V_{sd}$ being the voltage applied along the conducting channel of length $L$. The $(f(K)(1 - f(K')))$ represent the probability that the initial $k$ state is occupied and the final electron state $k'$ is empty while the factor $f(K')(1 - f(K))$ is that the initial $k$ state is empty and the final electron state $k'$ is occupied, in which $K' = \sqrt{k'^2 - 2k'k_d\cos(\theta + \phi) + k_d^2}$. In eq.(4), the summation over $k$ and $k'$ can be transformed into two dimensional integrals by $\sum_{k,k'} \to A^2/(2\pi)^4 \int dk dk'$. Taking $(N_q(t) + 1) = N_q(t)$ and neglecting $f(K)f(K')$, eq.(4) transforms to,

$$\frac{\partial N_q(t)}{\partial t} = (2\pi/\hbar)\, g_s g_v \left| M_{kk'}^{DP(PE)}(q) \right|^2 \int_0^\infty k\,dk \int_0^{2\pi} d\phi \int_0^\infty k'\,dk' \int_0^{2\pi} d\theta \left(\frac{1+\cos\theta}{2}\right) N_q(t) \times [f(K) - f(K')]\delta(E_k - E_{k'} - \hbar\omega_q) \quad (5)$$

Using eq.(5), $R_q^{DP/PE}$ can be expressed as,

$$R_q^{DP(PE)} = \frac{2\pi g_s g_v}{\hbar} \frac{A^2 \left| M_{kk'}^{DP(PE)}(q) \right|^2}{(2\pi)^4 \hbar v_f} \int_0^\infty k\,dk \int_0^{2\pi} d\phi \int_0^\infty k'\,dk' \int_0^{2\pi} d\theta \left(\frac{1+\cos\theta}{2}\right) \left\{ \frac{1}{1+\exp\left[\beta\hbar v_f \sqrt{k^2 - 2kk_d\cos\phi + k_d^2}\right]} - \frac{1}{1+\exp\left[\beta\hbar v_f \sqrt{k'^2 - 2k'k_d\cos(\theta+\phi) + k_d^2}\right]} \right\} \times \delta\left(k - k' - \frac{\omega_q}{v_f}\right) \quad (6)$$

A closed form solution of the eq.(6) is not possible because integration over $\theta$ and $\phi$ makes it non trivial, and we defer its numerical solution to section 3. To obtain an analytical solution we evaluate the integral under the simplifying assumption that the electrons and phonons are propagating along the direction of the applied field, that is we make $\theta$ and $\phi = 0$. Therefore the eq.(6) takes the form

$$R_q^{DP(PE)} = \frac{\pi g_s g_v}{\hbar} \frac{A^2 \left| M_{kk'}^{DP(PE)}(q) \right|^2}{(2\pi)^4 \hbar v_f} \int_0^\infty k\,dk \int_0^{2\pi} d\phi \int_0^\infty k'\,dk' \int_0^{2\pi} d\theta\, \delta\left(k - k' - \frac{\omega_q}{v_f}\right)$$

$$\times \left\{ \frac{1}{1+e^{[\beta\hbar v_f(k-k_d-k_f)]}} - \frac{1}{1+\exp[\beta\hbar v_f(k-\frac{\omega_q}{v_f}-k_d-k_f)]} \right\} \quad (7)$$

Both the attenuation coefficients for DP and PE interaction can be evaluated from eq.(7). The same explicitly for the case of DP coupling reduces to the following equation,

$$R_q^{DP} = \frac{AV^{DP^2}\omega_q}{2\pi\hbar\rho v_f v_s^{DP^2}} \int_0^\infty k\left(k-\frac{\omega_q}{v_f}\right)\left(\frac{1}{1+\exp[\beta\hbar v_f(k-k_d-k_f)]} - \frac{1}{1+\exp[\beta\hbar v_f(k-\frac{\omega_q}{v_f}-k_d-k_f)]}\right)dk \quad (8)$$

The above equation on integration yields the following solution,

$$R_q^{DP} = \frac{AV^{DP^2}\omega_q}{2\pi\hbar^4\rho\beta^3 v_s^{DP^2} v_f^4}(\beta\omega_q\hbar(\text{PolyLog}[2,-e^{(k_d+k_f)v_f\beta\hbar}] - \text{PolyLog}[2,-e^{\beta((k_d+k_f)v_f+\omega_q)\hbar}]) - 2\text{PolyLog}[3,-e^{(k_d+k_f)v_f\beta\hbar}] + 2\text{PolyLog}[3,-e^{\beta((k_d+k_f)v_f+\omega_q)\hbar}]) \quad (9)$$

Similarly on feeding the matrix element for PE interaction in eq.(7), yields the equation as under,

$$R_q^{PE} = \frac{Ac_{PE}^2 V^{PE^2} e^{-2qd}}{64^2\hbar\rho_s v_f v_s^{PE}} \int_0^\infty k\left(k-\frac{\omega_q}{v_f}\right)\left(\frac{1}{1+\exp[\beta\hbar v_f(k-k_d-k_f)]} - \frac{1}{1+\exp[\beta\hbar v_f(k-\frac{\omega_q}{v_f}-k_d-k_f)]}\right)dk \quad (10)$$

Which on integration gives the result,

$$R_q^{PE} = \frac{Ac_{PE}^2 V^{PE^2} e^{-2qd}}{64\pi^2\hbar^4\rho_s\beta^3 v_f^4 v_s^{PE}}(\beta\omega_q\hbar(\text{PolyLog}[2,-e^{(k_d+k_f)v_f\beta\hbar}] - \text{PolyLog}[2,-e^{\beta((k_d+k_f)v_f+\omega_q)\hbar}]) - 2\text{PolyLog}[3,-e^{(k_d+k_f)v_f\beta\hbar}] + 2\text{PolyLog}[3,-e^{\beta((k_d+k_f)v_f+\omega_q)\hbar}]) \quad (11)$$

In previous works using the BTE the FD function has been approximated at low temperature as $f(k) = \exp(-\beta(\hbar v_f k - \varepsilon_f))$, using which the eq. (7) reduces to

$$R_q^{DP} = \frac{AV^{DP^2}q}{2\pi\hbar\rho v_f v_s^{DP}} \int_0^\infty k\left(k-\frac{\omega_q}{v_f}\right)e^{-\beta\hbar(kv_f+\omega_q-v_f(k_d+k_f))}(-1+e^{\beta\omega_q\hbar})\Theta\left(k-\frac{\omega_q}{v_f}\right)dk \quad (12)$$

The solution to the above equation as reported in [30] is

$$R_q^{DP} = \frac{AV^{DP^2}qk_f^2}{2\pi\hbar^2\rho\beta v_s^{DP^4} v_f^4} v_d\left(v_d - \frac{q}{k_f}v_s^{DP}\right)e^{-\beta\hbar k_f v_d}(1-e^{\beta\hbar\omega_q}) \quad (13)$$

And the same with a slight modification in assumption is reported in [31] as

$$R_q^{DP} = \frac{AV^{DP^2}\omega_q}{2\pi\hbar^4\rho\beta^3 v_s^{DP^2} v_f^4}\left(2-\beta\omega_q\hbar\left(1-\frac{v_d}{v_s^{DP}}\right)\right)e^{-\beta\hbar k_f v_d}\left(1-e^{-\beta\hbar\omega_q\left(1-\frac{v_d}{v_s^{DP}}\right)}\right) \quad (14)$$

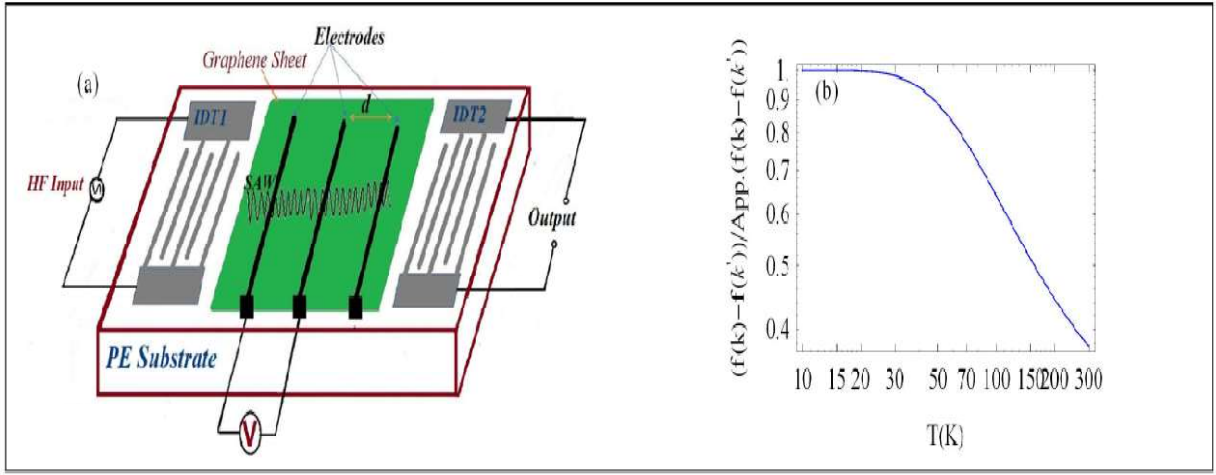

Fig.1 (a) Schematics of a Graphene sheet placed on a piezoelectric substrate between two IDT devices. A SAW is generated by a high-frequency signal input to IDT1 formed on a piezoelectric substrate which reaches IDT2 where it is converted back into high frequency signal for detection. When a SAW propagates in graphene, an acoustoelectric current ($I_{AE}$) flows between two attached electrodes.(b)Ratio of the difference between the complete and approximate Fermi-Dirac functions at $\omega_q = 1 THz$ and $k = 2 \times 10^8 m^{-1}$.

| Coefficient Name → <br> Result ↓ | Amplification coefficient, $R_q^{DP(PE)} (\sec^{-1})$ | Acoustoelectric current density, $J_{AE}^{DP(PE)} (Am^{-2})$ |
|---|---|---|
| Case of electron DP coupling | $\frac{AV^{DP^2}\omega_q K_B{}^3 T^3}{\pi\hbar^4 \rho v_s^{DP^2} v_f{}^4}(\frac{\omega_q\hbar}{K_B T}$ <br> $\times (\text{PolyLog}[2,-e^{(k_d+k_f)v_f\beta\hbar}]$ <br> $- \text{PolyLog}[2,-e^{\beta((k_d+k_f)v_f+\omega_q)\hbar}])$ <br> $- 2\text{PolyLog}[3,-e^{(k_d+k_f)v_f\beta\hbar}]$ <br> $+ 2\text{PolyLog}[3,-e^{\beta((k_d+k_f)v_f+\omega_q)\hbar}])$ | $\frac{-eV^{DP^2}\omega_q K_B{}^3 T^3}{\pi\hbar^4 \rho v_s^{DP^2} v_f{}^4}(\frac{\omega_q\hbar}{K_B T} \times (\text{PolyLog}[2,-e^{(k_d+k_f)v_f\beta\hbar}] -$ <br> $\text{PolyLog}[2,-e^{\beta((k_d+k_f)v_f+\omega_q)\hbar}]) -$ <br> $2\text{PolyLog}[3,-e^{(k_d+k_f)v_f\beta\hbar}] +$ <br> $2\text{PolyLog}[3,-e^{\beta((k_d+k_f)v_f+\omega_q)\hbar}])$ |
| Case of electron PE coupling | $\frac{Ac_{PE}^2 V^{PE^2} e^{-2qd}}{64\pi^2\hbar^4 \rho_s \beta^3 v_f{}^4 v_s^{PE}}(\beta\omega_q\hbar(\text{PolyLog}[2,-e^{(k_d+k_f)}$ <br> $- \text{PolyLog}[2,-e^{\beta((k_d+k_f)v_f+\omega_q)\hbar}])$ <br> $- 2\text{PolyLog}[3,-e^{(k_d+k_f)v_f\beta\hbar}]$ <br> $+ 2\text{PolyLog}[3,-e^{\beta((k_d+k_f)v_f+\omega_q)\hbar}])$ | $\frac{-e\, c_{PE}^2 V^{PE^2} e^{-2qd}}{64\pi^2\hbar^4 \rho_s \beta^3 v_f{}^4 v_s^{PE}}\left(\beta\omega_q\hbar\left(\text{PolyLog}[2,-e^{(k_d+k_f)v_f\beta}]\right.\right.$ <br> $- \text{PolyLog}\left[2,-e^{\beta((k_d+k_f)v_f+\omega_q)\hbar}\right]\Big)$ <br> $- 2\text{PolyLog}\left[3,-e^{(k_d+k_f)v_f\beta\hbar}\right]$ <br> $+ 2\text{PolyLog}\left[3,-e^{\beta((k_d+k_f)v_f+\omega_q)\hbar}\right]\Big)$ |
| Earlier Reported Result in BTF for electron DP coupling in BG Regime | $\frac{AV^{DP^2} q k_f^2 K_B T}{2\pi\hbar^2 \rho v_s^{DP} v_f{}^4} v_d\left(v_d - \frac{q}{k_f}v_s^{DP}\right)e^{-\beta\hbar k_f v_d}$ <br> $(1 - e^{\beta\hbar\omega_q})$ | $-\frac{16\,\tau\, qV^{DP^2} v_s^{DP^3} T^4}{\pi\,\rho\,\hbar^3 T_{BG}{}^4}(2-\beta\omega_q\hbar)(1-e^{-\beta\omega_q\hbar})$ |

**Table1**. Summary of the obtained and reported analytical expressions for Amplification coefficient, $R_q^{DP(PE)}$ and acoustoelectric current density, $J_{AE}^{DP(PE)}$ for the cases of electron acoustic phonon coupling via Deformation potential and Piezoelectric modes.

The $J_{AE}^{DP(PE)}$ is directly proportional to the acoustic phonon population rate as $J_{AE}^{DP(PE)} \propto R_q^{DP(PE)}$. However, the expression for $J_{AE}^{DP(PE)}$ in MLG quoted in Ref.[31] as $J_{AE}^{DP(PE)} = \frac{2e\tau}{\hbar v_f} \Gamma_q^{DP(PE)}$, where $\tau$ is the relaxation time and $\Gamma_q^{DP(PE)} = v_s^{DP(PE)} R_q^{DP(PE)}$, is suspect as the expression does not bears the dimension of current density. Another correct phenomenological expression obtained in the classical relaxation model is $J_{AE}^{DP(PE)} = \frac{-\mu I}{v_s^{DP(PE)}} \Gamma_q^{DP(PE)}$, in which $\mu$ is the mobility and $I$ is the intensity of the SAW wave. Since the prefactors in the above expression, i.e., $\frac{-\mu I}{v_s^{DP(PE)}}$ have the dimension of linear current density (current/length) therefore we obtain an empirical relation for current density by multiplying $R_q^{DP(PE)}$ with $\frac{e}{A}$. Hence, $J_{AE}^{DP(PE)} = -\frac{eR_q^{DP(PE)}}{A}$, which on substituting the full expression for $R_q^{DP(PE)}$ from eq.(6) becomes $J_{AE}^{DP(PE)} = -\frac{e}{A} \frac{1}{N_q} \frac{\partial N_q}{\partial t}$. Therefore,

$$J_{AE}^{DP(PE)} =$$

$$\frac{2\pi g_s g_v}{\hbar} \frac{Ae|M_{kk'}^{DP(PE)}(q)|^2}{(2\pi)^4 \hbar v_f} \int_0^\infty k\,dk \int_0^{2\pi} d\phi \int_0^\infty k'\,dk' \int_0^{2\pi} d\theta \left(\frac{1+\cos}{2}\right) \left\{ \frac{1}{1+\exp\left[\beta \hbar v_f \sqrt{k^2 - 2kk_d \cos\phi + k_d^2}\right]} - \frac{1}{1+ex\left[\beta \hbar v_f \sqrt{k'^2 - 2k'k_d \cos(\theta+\phi) + k_d^2}\right]} \right\} \times \delta\left(k - k' - \frac{\omega_q}{v_f}\right) \quad (15)$$

Feeding the obtained expressions, eqs.(8) and (9) for $R_q^{DP(PE)}$, derived in the approximation that electrons and phonons are propagating along the direction of the applied field, that is $\theta$ and $\phi = 0$, in $J_{AE}^{DP(PE)}$ we get the following separate expressions for $J_{AE}^{DP}, J_{AE}^{PE}$, respectively as,

$$J_{AE}^{DP} = \frac{-eV^{DP^2}\omega_q}{2\pi\hbar^4 \rho \beta^3 v_s^{DP^2} v_f^4} (\beta\omega_q \hbar (\text{PolyLog}[2, -e^{(k_d+k_f)v_f \beta \hbar}] - \text{PolyLog}[2, -e^{\beta((k_d+k_f)v_f+\omega_q)\hbar}]) -$$
$$2\text{PolyLog}[3, -e^{(k_d+k_f)v_f \beta \hbar}] + 2\text{PolyLog}[3, -e^{\beta((k_d+k_f)v_f+\omega_q)\hbar}]) \quad (16)$$

$$J_{AE}^{PE} =$$
$$\frac{-ec_{PE}^2 V^{PE^2} e^{-2qd}}{64\pi^2 \hbar^4 \rho_s \beta^3 v_s^{PE}} \left(\beta\omega_q \hbar \left(\text{PolyLog}[2, -e^{(k_d+k_f)v_f \beta \hbar}] - \text{PolyLog}\left[2, -e^{\beta\left((k_d+k_f)v_f+\omega_q\right)\hbar}\right]\right) -$$
$$2\text{PolyLog}[3, -e^{(k_d+k_f)v_f \beta \hbar}] + 2\text{PolyLog}\left[3, -e^{\beta\left((k_d+k_f)v_f+\omega_q\right)\hbar}\right]\right) \quad (17)$$

Eqns.(16) and (17) respectively represent the complete unscreened AE $J_{AE}^{DP(PE)}$ through coupling with DP and PE phonons in MLG.

The same has been calculated in Ref. [31] in low temperature Bloch-Gruneisen (BG) regime as under,

$$J_{AE}^{DP} = j_0 \{2 - \beta\omega_q\hbar\}[1 - \exp(-\beta\omega_q\hbar)] \quad (18)$$

where $j_0 = -16\tau q V^{DP^2} v_s^{DP^3} T^4 / \pi\rho\hbar^3 T_{BG}^4$, where $T_{BG} = 2\hbar v_s^{DP} k_f / k_B$.

The AE current can also be calculated using the eq.(13) as

$$J_{AE}^{DP} = \frac{eV^{DP^2} q k_f^2}{2\pi\hbar^2 \rho\beta} v_d \left(v_d - \frac{q}{k_f} v_s^{DP}\right) e^{-\beta\hbar k_f v_d} \left(1 - e^{\beta\hbar\omega_q}\right) \quad (19)$$

### III. NUMERICAL AND ANALYTICAL RESULTS AND DISCUSSIONS

In the following the numerical evaluation of the complete expression and the obtained analytical results is carried out. We investigate the role of DP and PE coupling on the generation of AE current as a function of temperature, frequency, phonon velocity and electronic density. For the estimation of numerical results we use the following values of parameters: $V^{DP} = 9$eV, $v_s^{DP} = 2.1 \times 10^3$m/sec, $\rho = 6.5 \times 10^{-7}$ kg/m$^2$, $\tau = 10^{-10}$sec, $A = 200 \times 100 \ \mu m^2$, $\omega_q = 1.5 \times 10^{12} s^{-1}$, $V^{PE} = 2.4 \times 10^7$ eV/cm, $\rho_s = 5.3$g/cm$^3$, $c_{PE} = 4.9$ (known by the elastic properties of GaAs), $v_s^{PE} \approx 0.9 v_b \approx 2.7 \times 10^3 m/s$ [30-31]. The electron-PE acoustic phonon coupling is strongest for surface phonons propagating along the diagonal direction with $q_x \approx q_y$, so we can approximate $\frac{q_x q_y}{q^2} \approx \frac{1}{4}$, for large distances $d > 5$Å, so $e^{-2qd} \approx 1$.

It was pointed out in the introduction that the approximations made in obtaining the earlier reported analytical result stands correct in the low temperature limit $(T \leq 30K)$, as then the numerical factor of 1 in comparison to exponential factor in the denominator of the FD function, $f(K) = \{\exp[\beta(\hbar v_f K - \varepsilon_f)] + 1\}^{-1}$ is small, but for higher temperatures the same is not true as the term $\exp[\beta(\hbar v_f K - \mu)]$ begins to decline fast with rising temperature, $(T > 30K)$. The same can be clearly observed from the fig.1b where a plot of ratio of the difference between the complete and approximate FD functions, $f(K) - f(K')/Appr.(f(K) - f(K'))$ at $\omega_q = 1 THz$ and $k = 2 \times 10^8 m^{-1}$ is depicted. Clearly the ratio is 1 in the low temperature limit and above $(T > 30K)$ the ratio begins to decline. So the reported results under this approximation at higher temperatures may not represent the accurate picture and hence are suspect. Next we consider the numerical estimation of the obtained analytical results for AE current for DP and PE phonon coupling and their comparison with full numerical computation of the governing eq.(15). In fig.2.(a) we plot the less restrictive approximate analytical result from eq.(16) for three different temperatures T=95K, 200K, 300K at $v_s^{DP} = 2.1 \times 10^3 ms^{-1}$. The magnitude of current at $\omega_q = 2.7 \times 10^8 Hz$ is about 18(nA) at T=300K. We find that there is a slight decrease in the current with increase in temperature which further increases with increasing frequency. The fig.2(b) gives a comparative plot of the current, $I_{AE}$ vs $\omega_q$ at T=100K at $v_s^{DP} = 2.1 \times 10^4 ms^{-1}$, where curve-A, corresponds to the full numerical result using eq.(15), curve-B from eq.(16), and curve-C using eq.(14) after feeding it into the definition of $I_{AE}^{DP} = -eR_q^{DP}$. The curve-B is 20 times more in magnitude than curve-C at $2 \times 10^{11}$Hz. Clearly the reported analytical result underestimates the amplitude coefficient. Overall from fig. 2a, the order of the obtained magnitude of the current in nA for frequencies in hundreds of MegaHertz [22], and in mA from fig.2b are in excellent agreement with the reported values.

The fig.3(a) depicts the approximate analytical result for $I_{AE}^{DP}$ from eqs.(16) & (13) respectively up to 0.2 GHz at T=20K & $v_s^{DP} = 2.1 \times 10^3 ms^{-1}$. The temperature dependence of the same is plotted in fig.3(b). The loglog plot of the same with temperature at $\omega_q = 1THz$ & $v_s^{DP} = 2.1 \times 10^4 ms^{-1}$, where curve-A is from eq.(19), curve-B from eq.(16). The curve-A from earlier reported result eq.(16) has weaker temperature dependence at higher frequencies while our obtained result incorporating the full FD functions exhibits a stronger temperature dependence at lower temperatures (<100K).

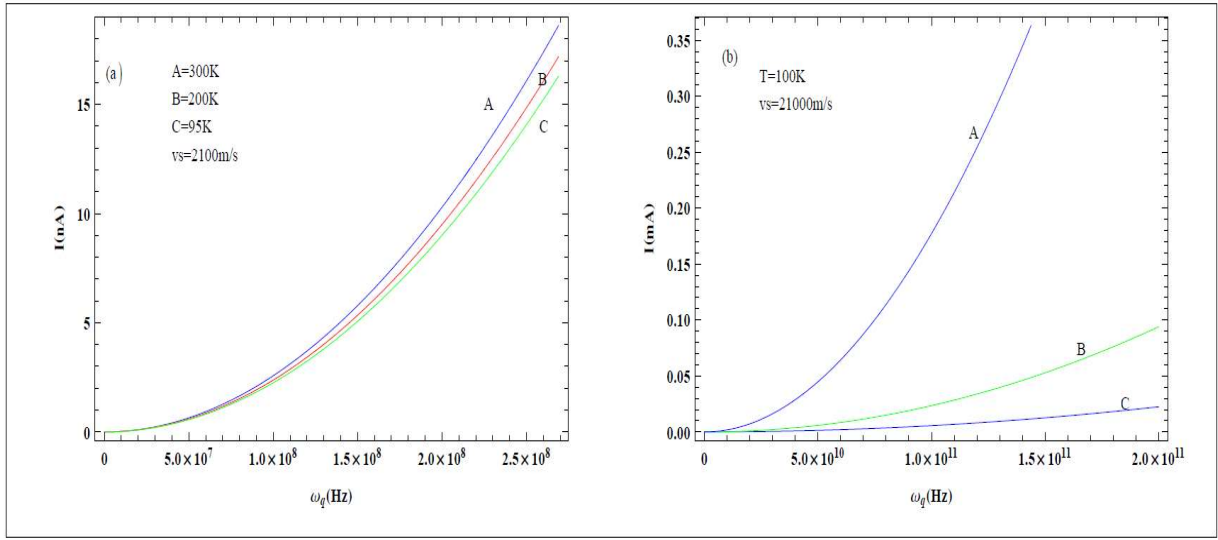

Fig.2. (a) $I_{AE}$(nA) vs $\omega_q$(Hz) plot from the analytical result from eq.(16) for three different temperatures (curve-A at 300K, curve-B at 200K and curve-C at 95K), at $v_s^{DP} = 2.1 \times 10^3 ms^{-1}$. (b) Comparative plot of the current $I_{AE}$(mA) vs $\omega_q$(Hz) at T=100K and $v_s^{DP} = 2.1 \times 10^4 ms^{-1}$, where curve-A, is full numerical from eq.(15), curve-B from eq.(16) and curve-C from eq.(19).

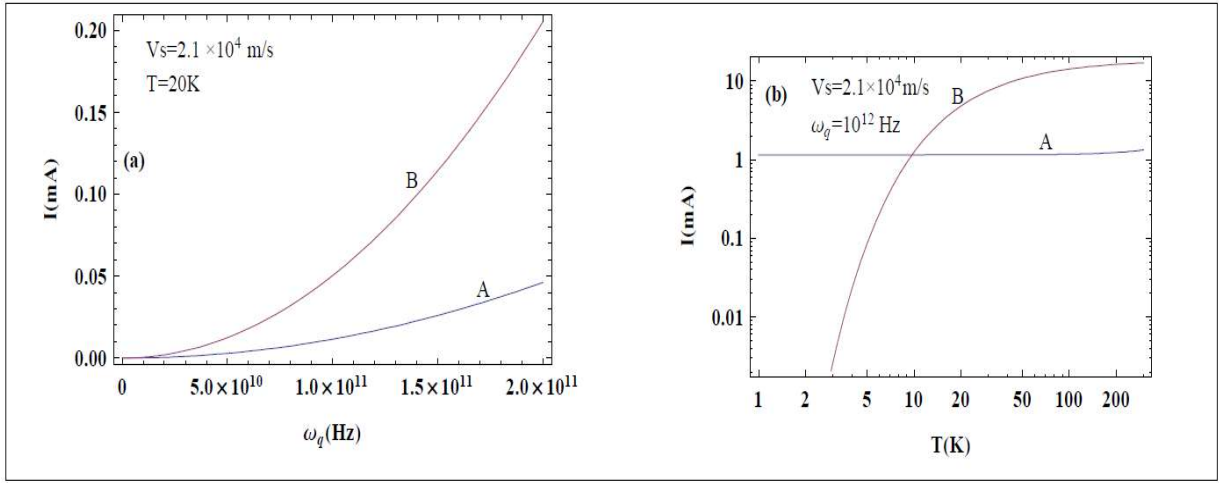

Fig.3. (a) $I_{AE}$(mA) vs $\omega_q$(Hz) plot from approximate analytical results curves-A and B given by eqs.(19) &(16), respectively at, T=20K & $v_s^{DP} = 2.1 \times 10^3 ms^{-1}$. (b) Loglogplot of $I_{AE}$(mA) vs $T$(K) of the same at $\omega_q = 1\text{THz} \& v_s^{DP} = 2.1 \times 10^4 ms^{-1}$, where curve-A, is from eq.(19), curve-B from eq.(16).

Besides frequency and temperature the $I_{AE}$ also depends on phonon velocity and carrier density. To investigate the dependence of AE current with velocity and carrier density we plot in fig.4(a) the approximate analytical results from eqs.(16) &(19), respectively as a function of $v_s^{DP}$ at T=100K & $\omega_q = 1\text{THz}$. And in fig.4(b) the ratio of the currents from eqs.(16) and (19) as a function carrier concentration at $\omega_q = 1\text{THz}$, T=100K is shown. From fig 4(a) we notice that the two curves A and B from eqs.(16) and (19), respectively, show a contrasting behavior with phonon velocity. The curve-A from our obtained expression declines with increasing velocity while the curve-B from the earlier reported result increases with phonon velocity. The same type of behavior from the two results is seen also seen with carrier density as plotted in fig5(b). Initially the curve-A from eq.(16) dominates but after a cross over near $3.53.5 \times 10^{15} cm^{-2}$, the curve –B begins to dominate.

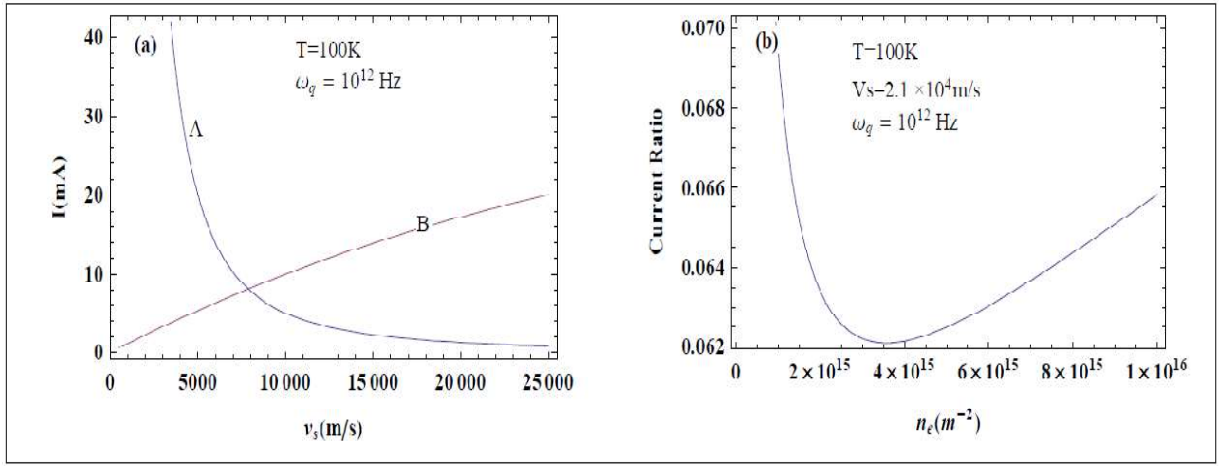

Fig.4.(a) Approximate analytical result from eqs.(16) &(19) respectively as a function of $v_s^{DP}(m/s$ at T=100K & $\omega_q = 1\text{THz}$. (b) Ratio of the currents from eq.(16)/eq.(19) as a function carrier concentrarion at $\omega_q = 1\text{THz}$ and T=100K.

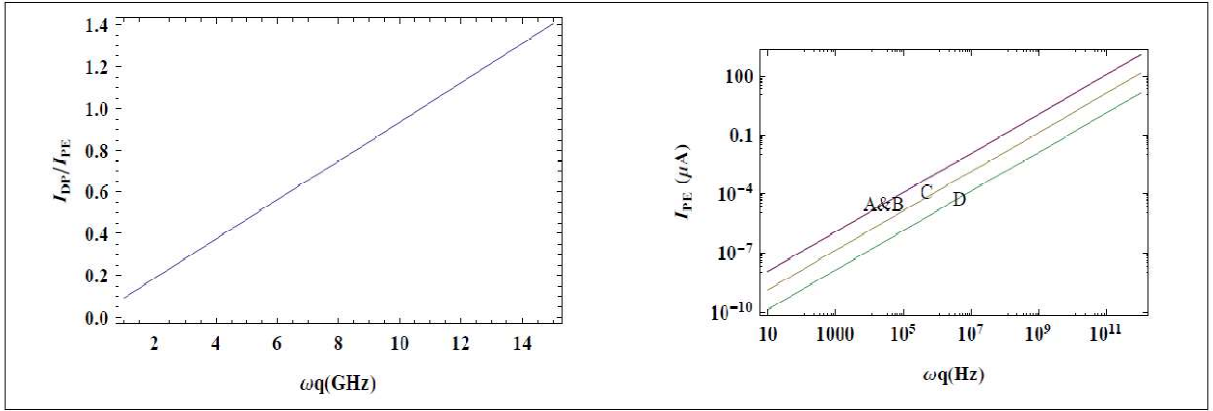

Fig5. (a) Ratio of the currents $I_{DP}/I_{PE}$ with $\omega_q$(GHz). (b) Analytical $I_{PE}$ with $\omega_q$(Hz), Top merging curves A & B are at $T = 20K$ and 100K respectively, for $n_e = 1 \times 10^{16} cm^{-2}$ and $v_S^{PE} = 2.7 \times 10^3 m/s$, middle curve-C at $T = 100K$ and $n_e = 1 \times 10^{15} cm^{-2}, v_S^{PE} = 2.7 \times 10^3 m/s$, lowest curve-D at $T = 100K$ and $n_e = 1 \times 10^{15} cm^{-2}, v_S^{PE} = 2.7 \times 10^4 m/s$.

To check the relative significance of PE and DP coupling currents we plot the ratio of $I_{DP}/I_{PE}$ employing eqs. (16) & (17) as a function of frequency $\omega_q$(GHz) in fig.5a. This ratio is independent of temperature, and near to 11 GHz the ratio is equal. So we can infer that the PE contribution remains dominant below 11 G Hz and above it the DP forms the major source of acoustoelectric current. In fig.5b is plotted the analytical $I_{PE}$ vs $\omega_q$(GHz) for different set of carrier densities, phonon velocities and temperatures. The analytical $I_{PE}$ with $\omega_q$(Hz), top merging curves-A & B are at $T = 20K$ and 100K respectively, for $n_e = 1 \times 10^{16} cm^{-2}$ and $v_S^{PE} = 2.7 \times 10^3 m/s$, middle curve-C at $T = 100K$ and $n_e = 1 \times 10^{15} cm^{-2}, v_S^{PE} = 2.7 \times 10^3 m/s$, lowest curve-D at $T = 100K$ and $n_e = 1 \times 10^{15} cm^{-2}, v_S^{PE} = 2.7 \times 10^4 m/s$. From this figure we observe that by increasing temperature $T = 20K$ and 100K keeping other variables fixed the current remains almost the same. But the current has stronger dependence on carrier density and phonon velocity as it decreases on decreasing the carrier density and increasing the phonon velocity.

## IV. CONCLUSIONS

We investigated both numerically and analytically the governing kinetic equations on amplification/attenuation coefficient and acoustoelectric current in monolayer graphene due to piezoelectric and deformation potential electron phonon couplings in the Boltzmann transport formalism approach, by overcoming the earlier simplifying assumptions covering the low and high temperature range, and obtain analytical and numerical results. We also investigated the role of temperature, frequency, phonon velocity and electronic density besides the effect of DP and PE coupling on AE current. Our results also differ from the earlier reported results in the following sense; (i) the obtained results show a much stronger dependence (about 20 times) on frequency, (ii) it shows a relatively stronger dependence on temperature at higher frequencies, (iii) and with phonon velocity and carrier density as it shows a contrasting behavior with the earlier reported result, (iv) at higher frequencies the DP contribution becomes significant and above 11 G Hz the DP dominates the PE contribution and forms the major source of acoustoelectric current. We find that the results obtained are in reasonable conformity with the reported experimental results. Further the obtained magnitude of acousto electric current for very high frequencies ($10^{12}$Hz) frequencies at room temperatures augurs well for graphene to be used as a very high frequency SAW device.